\documentclass[apj]{emulateapj}
\usepackage{apjfonts}
\usepackage{natbib}
\usepackage{graphicx}
\bibpunct{(}{)}{;}{a}{}{,}

\begin{document}
   \title{A simple model for the distribution of
   quiet Sun magnetic field strengths}
   \author{
   	J.~S\'anchez~Almeida,
	}
    \affil{Instituto de Astrof\'\i sica de Canarias, 
              E-38205 La Laguna, Tenerife, Spain}
   \email{jos@iac.es}

\begin{abstract}
We derive a first order linear differential equation
describing the shape of
the probability density function  of 
magnetic field strengths in the quiet Sun (PDF).
The modeling is very schematic. It considers 
convective motions which continuously supply 
and withdraw magnetic structures. In addition, a 
magnetic amplification mechanism increases the field 
strength up to a threshold that cannot be exceeded. 
These three basic ingredients provide PDFs in  good agreement 
with the PDFs produced by
realistic numerical simulations of magneto convection,
as well as with quiet Sun PDFs inferred from observations.
In particular, the distribution is approximately lognormal, 
and it produces an excess of
magnetic fields (i.e., a {\em hump} in the distribution) 
right before the maximum field strength. 
The success of this simple model may indicate that
only a few basic ingredients shape the quiet Sun PDF. 
Our approach provides a concise parametric 
representation 
of the PDF, as required to develop automatic methods of 
diagnostics.
\end{abstract}

%
\maketitle

\section{Introduction}\label{introduction}

\citet{liv75} and \citet{smi75} discovered weak 
polarization signals in the interior of  supergranulation 
cells. These magnetic signals are known as Inter-Network 
magnetic fields (IN),  Intra-Network fields or, simply, quiet 
Sun fields.
It has been long conjectured that such signals
trace a hidden component of the solar 
magnetic field having most of the 
unsigned magnetic flux and magnetic 
energy \citep[e.g.,][]{unn59,
ste82,yi93,san98c,san04,sch03b}.
This suggestion seems to be confirmed by recent 
measurements \citep[e.g.][]{ste82,fau93,fau95,san00,soc02,
tru04,man04,san05}, as well as by numerical simulations
of magneto convection 
\citep[e.g.][]{cat99a,emo01,ste02,vog03b,vog05}.
The signals discovered by \citeauthor{liv75} and
\citeauthor{smi75} represent the residual left
when a magnetic field of complex topology is observed
with finite angular resolution \citep[e.g.,][]{emo01,san03}. 
Such complex
field is to be expected as a result of the MHD interaction
between magnetic fields 
and the random motions associated  with the  
granulation.

\begin{figure}
\includegraphics[width=0.45\textwidth]{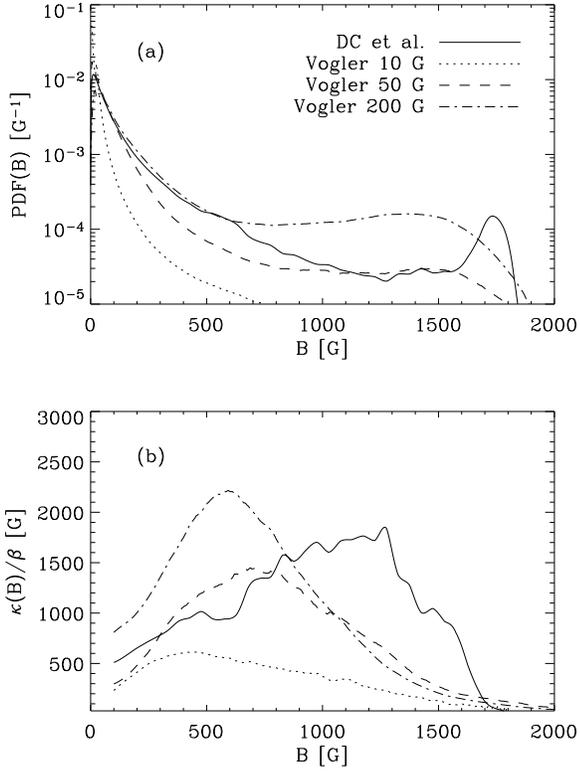}
\caption{(a) Set of quiet Sun PDFs derived from numerical 
simulations of magneto convection (the discontinuous lines),
and a semi-empirical PDF that fits observed Zeeman and 
Hanle signals (the solid line). (b) $\kappa$s 
reproducing the PDFs shown above. 
The labels in the inset stand for:
 Vogler $\equiv$ \citet{vog03b}, and
DC~et~al. $\equiv$ \citet{dom06}.
%
}
\label{pdf}
\end{figure}

Characterizing these IN fields is therefore important, and 
significant advances have been produced during the last 
years (see the references given above).
Among the observational parameters used for characterization,
the probability density function  of magnetic field
strengths (PDF) turns out to be particularly useful. 
It is defined as the fraction of quiet Sun occupied by 
magnetic fields of each strength. It condenses basic
physical information like the filling factor,
the unsigned flux, and the magnetic energy 
corresponding to each field strength 
\citep[see ][]{san04}. 
In addition, some of the diagnostic 
techniques employ them directly 
\citep[e.g.,][]{fau93,fau95,fau01,tru04}, 
and PDFs are predicted by the numerical 
simulations, allowing a direct comparison between 
simulations and observations 
\citep{cat99a,ste02,vog03,vog05}.  
In spite of these advantages, 
measuring the quiet Sun PDF is not trivial.
All direct estimates are strongly 
biased. The Zeeman signals originate only in 
regions of significant field strength and net
polarity, whereas 
the Hanle signals are insensitive to the 
fields stronger than a few hundred~G.   
\citet{dom06} carry out a first attempt to 
provide the full PDF from 0~G to 1800~G. They
remove the known biases by using Hanle effect
measurements to constrain the shape at weak fields 
($<$~200~G), Zeeman signals for the strong fields 
(say, $>$~400~G), and assuming the PDF to be 
continuous in between.
They work out a set of PDFs compatible
with the observations and consistent with 
numerical simulations of magneto-convection.
As it is acknowledged by the authors, they carry out
an exploratory estimate and therefore further independent
work is required for confirmation.
Support would be provided if some 
of the characteristic and  unexpected properties 
of these PDFs can be explained in terms of simple
physical mechanisms. In particular, the  
empirical PDFs show an increase of strong 
kG fields right before the maximum possible field 
strength, set by the gas pressure of the quiet photosphere
(see the solid line in Fig.~\ref{pdf}a with a
 {\em hump} at $B\sim 1700$~G). 
Despite the fact that these strong fields 
occupy only a small fraction of the surface, 
they contribute with a large part of the
quiet Sun unsigned magnetic flux and magnetic
energy, which makes them particularly 
important. Moreover the kG fields 
turn out to be ideal for studying the quiet Sun 
magnetism since they show up in unpolarized 
light images \citep{san04a}.
The possible physical origin of such a 
hump is addressed here. 
We discuss that a terminal hump is to be
expected if a magnetic amplification 
mechanism operates in the quiet Sun.
These mechanisms have a long tradition to explain
the presence of kG magnetic fields in plage and
network regions 
\citep{wei66,par78,spr79,van84,san01c,cam05}.
They concentrate weakly magnetized plasma 
conserving the magnetic flux, which demands 
increasing the field strength.  No matter the details 
of the mechanism, all of them abruptly die out 
at the  maximum field strength imposed by the
gas pressure of the quiet photosphere. The magnetic 
structures 
tend  to pile up at this limit giving rise to a hump 
(bottleneck effect). The idea was already put forward by 
\citet{dom06b}, but it is elaborated in this paper. 

We write down a differential
equation to describe the time evolution of the IN
PDF (\S~\ref{setup}). The stationary solutions of 
such equation are analyzed in 
\S~\ref{properties}. 
It is shown  how a terminal hump appears in 
a natural way. 
The efficiency of the amplification mechanism is 
parameterized using a coefficient (speed) with units of 
magnetic field strength per unit time. 
Given a PDF, one can infer by inversion
the speed of the concentration mechanism. 
The technique is applied to PDFs coming from 
numerical simulations, as well as to
 semi-empirical PDFs (\S~\ref{selected}).
Other uses and limitations 
of the model PDFs are discussed in 
\S~\ref{discussion}.


%
\section{Differential Equation}\label{setup}

We use a heuristic approach to derive an 
equation for the shape of the PDF $P(B)$.
The different mechanisms creating, destroying 
and modifying magnetic 
fields determine $P(B)$.   Here we
consider three of them plus the
buffeting of the granulation,
namely, (1) a magnetic  
amplification mechanism that tends to increase the magnetic 
field strength, 
(2) the submergence of the existing flux transported by 
the granulation downdrafts, (3) the emergence of new 
magnetic flux transported by the granulation upflows,
and (4) the buffeting of the granular motions 
on the magnetic structures. 
In the stationary state they are perfectly 
balanced, therefore,
\begin{equation}
{{dP(B)}\over{dt}}={{dP_1(B)}\over{dt}}+
	{{dP_2(B)}\over{dt}}+{{dP_3(B)}\over{dt}}
	+{{dP_4(B)}\over{dt}}=0, 
\label{eq0}
\end{equation}
where $dP_j/dt$ represents the variation with time
produced by the $j$-th mechanism. We proceed by
working out the contribution of the four terms
separately, and then considering all them 
together.
Simplicity is the main driver therefore a number
of simplifying hypotheses restrict the problem 
to make it tractable.

\begin{figure}
\includegraphics[width=.28\textwidth]{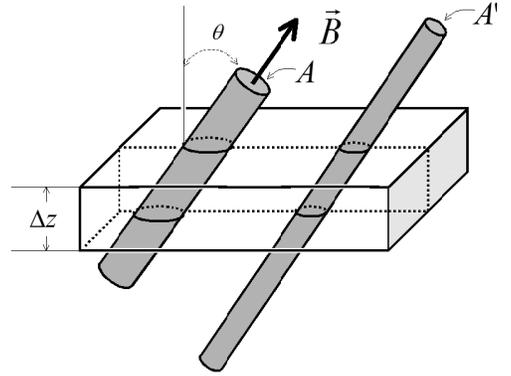}
\caption{
Schematic with a portion of the 
photosphere (the box) which includes the full range of 
photospheric heights $\Delta z$.
The two cylinders represent magnetic fluxtubes. 
The photospheric volume occupied by one of such 
fluxtubes depends on its cross-section  
$A$ and its inclination $\theta$, but not
on its length if it is longer than $\Delta z/\cos\theta$. 
}
\label{cartoon1}
\end{figure}

First, consider the magnetic amplification mechanism.
We start off by assuming that the photospheric 
magnetic fields are made of fluxtubes, each one 
having a single magnetic field vector and a single
cross-section. Four parameters characterize each 
fluxtube; the magnetic field strength $B_i$, 
the magnetic field  inclination $\theta_i$, 
the magnetic field azimuth $\phi_i$, and the section $A_i$.
Then the properties  of the photospheric magnetic fields 
are fully characterized given the four parameters
$B_i,\theta_i, \phi_i$, and $A_i$ corresponding to each
fluxtube -- the subscript $i$ varies from one to the number
of fluxtubes present in the photosphere.  
By definition, and  when $\Delta B\rightarrow 0$,
$P(B)\,\Delta B$ is the fraction of 
photospheric volume occupied by all magnetic 
structures with strengths between $B-\Delta B/2$ and 
$B+\Delta B/2$.
It can be expressed as a sum that considers all fluxtubes,
\begin{equation}
P(B)\,\Delta B={1\over{\sum_i\,\upsilon_i}}
\sum_i\,\upsilon_i\,\Pi\Big({{B_i-B}\over{\Delta B}}\Big),
\label{rectangle}
\end{equation}
with $\upsilon_i$ the volume
of the $i-$th fluxtube, and $\Pi$ the rectangle
function,
\begin{equation}
\Pi(x)=\cases{1 &$|x| < 1/2$,\cr
	     0& elsewhere.}
\end{equation}
Given a field strength $B$, the sum in the 
numerator of equation~(\ref{rectangle}) differs from
zero only when $B-\Delta B/2 < B_i < B+\Delta B/2$ and,
therefore, it gives the total volume occupied by fluxtubes
with field strengths between $B-\Delta B/2$ and 
$B+\Delta B/2$. Note also that the normalization
constant in the denominator of equation~(\ref{rectangle})
is the volume of the photosphere and, therefore, 
a constant.
In the case of interest, when $\Delta B\rightarrow 0$,
the rectangle function can be replaced with a
Dirac $\delta$-function since,
\begin{equation}
\delta(x)=\lim_{\Delta x\rightarrow 0} 
{{\Pi(x/\Delta x)}\over{\Delta x}};
\end{equation}
see, e.g.,  \citet[][]{bra78}.
This replacement leads to the compact expression
for $P(B)$
adequate for mathematical manipulations,
\begin{equation}
P(B)={1\over{\sum_i\,\upsilon_i}}
\sum_i\,\upsilon_i\,\delta(B_i-B).
\label{rectangle2}
\end{equation}
Each fluxtube has a single inclination, therefore, we are 
implicitly assuming that the
fluxtubes are straight and longer than the range of heights
of the photosphere;
see Figure~\ref{cartoon1}.
Accordingly, the volume of photosphere occupied by a fluxtube 
is simply set by its section and inclination,
\begin{equation}
\upsilon_i=\Delta z\,A_i/\cos\theta_i,
	\label{area}
\end{equation}  
where $\Delta z$ stands for the vertical extent 
of the photosphere or,
if this is too large for 
the approximation~(\ref{area}) to hold,
$\Delta z$ represents 
the range of heights to be described by our PDF. 
We characterize the magnetic amplification mechanism 
with the rate $\kappa$ 
at which it changes the magnetic field strength,
\begin{equation}
\kappa(B_i)={{dB_i}\over {dt}}.
\label{def_speed}
\end{equation}
The rate has units of magnetic field strength per unit
time, and we will call it {\em speed}. 
According to equation~(\ref{def_speed}), the speed 
only depends on the field strength of the fluxtube.
This is a working hypothesis which may not hold in
some practical cases, as we point out
in \S~\ref{discussion}.
All the magnetic amplification mechanisms proposed so far
conserve the magnetic flux so that an increase of field
strength comes together with a decrease of the section of the
magnetic structure to maintain the product $A_i\,B_i$ 
constant\footnote{
It follows from  the  invariance of the magnetic flux 
under stretching of fluxtubes in perfectly conducting 
plasmas. See, e.g., \citet[][\S 1.2.3]{chi95}. 
},
\begin{equation}
{{d}\over {dt}}(B_i\,A_i)=0.
\label{flux1}
\end{equation}
Equations~(\ref{area}) and (\ref{flux1}) lead to,
\begin{equation}
{{d\upsilon_i}\over {dt}}=-{{\upsilon_i}\over{B_i}} \kappa(B_i),
\label{flux2}
\end{equation}
where we assume that the concentration mechanism
does not change the field inclination in a preferred 
sense.
Using equations (\ref{rectangle2}) and 
(\ref{flux2}), and the chain rule,
\begin{displaymath}
\Big(-\sum_i\,\upsilon_i\Big)\,{{dP_1(B)}\over{dt}}=
\end{displaymath}
\begin{equation}
\sum_i\,{{\kappa(B_i)}\over{B_i}}
\upsilon_i\,\delta(B_i-B)
+\sum_i\,\kappa(B_i)\,\upsilon_i\,
{{\delta(B_i-B)}\over{B_i-B}},
\label{mess1}
\end{equation} 
where we have employed the expression for the derivative of
a $\delta$-function \citep[e.g.,][]{bra78},
\begin{equation}
{{d\delta(x)}\over{dx}}=-\delta(x)/x.
\end{equation}
Keeping in mind the following properties,
\begin{equation}
{{\kappa(B_i)}\over{B_i}}
\delta(B_i-B)={{\kappa(B)}\over{B}}\,
\delta(B_i-B),
\end{equation}
and
\begin{equation}
\kappa(B_i)\,
{{\delta(B_i-B)}\over{B_i-B}}
\simeq 
\Big[\kappa(B)+{{d\kappa(B)}\over{dB}}(B_i-B)\Big]\,
{{\delta(B_i-B)}\over{B_i-B}},
   \label{mess2}
\end{equation}
one can rewrite equation~(\ref{mess2}) as, 
\begin{equation}
\kappa(B_i)\,
{{\delta(B_i-B)}\over{B_i-B}}
\simeq 
\kappa(B)\,
{{d\delta(B_i-B)}\over{dB}}+
{{d\kappa(B)}\over{dB}}\,\delta(B_i-B),
\end{equation}
and, consequently, equation~(\ref{mess1}) turns 
out to yield either,
\begin{equation}
-{{dP_1(B)}\over{dt}}\simeq 
B^{-1}\kappa(B)\,P(B)+\kappa(B){{dP(B)}\over{dB}}+
P(B)\,{{d\kappa(B)}\over{dB}},
\end{equation}
or, alternatively, 
\begin{equation}
-{{dP_1(B)}\over{dt}}\simeq{{1}\over{B}}
{{d[\kappa(B)\,B\,P(B)]}\over{dB}}.
\label{mess3}
\end{equation}
The previous equation describes the variation with time of 
the PDF due to a magnetic amplification mechanism 
with speed $\kappa(B)$. 
%
Note that if rather than an amplification mechanism
one deals with a magnetic {\em de-amplification}
mechanism, equation~(\ref{mess3}) remains valid
with $\kappa(B) < 0$.
Consequently, one can think of $\kappa(B)$ as the 
average between all the amplification and 
de-amplification mechanisms operating in the
quiet Sun.  The existence of de-amplifications
is indeed likely since they often come together
with the amplification mechanisms
\citep[e.g., ][]{spr79,cam05}.

Consider the magnetic fields transported
by the downdrafts. In this case magnetic
structures disappear in proportion to the
existing magnetic structures, i.e.,
\begin{equation}
{{dP_2(B)}\over{dt}}\simeq -\beta(B)\, P(B),
\label{eq2}
\end{equation}
where $\beta (B)$ provides the rate of submergence per
unit time. The rate $\beta(B)$ depends on $B$ to
consider that different photospheric field strengths
might submerge at different rates. 

The emergence of magnetic fields transported
by the upflows has the same structure of the submergence 
but considering  a PDF $M(B)$ not 
necessarily the same as the photospheric PDF. 
In this case,   
\begin{equation}
{{dP_3(B)}\over{dt}}\simeq \alpha(B)\, M(B),
\label{eq3}
\end{equation}
with the rate of emergence $\alpha(B)$
also depending on the field strength.

The interaction between granular motions and
magnetic fluxtubes changes magnetic field 
inclinations, azimuths, field strengths, and sections. 
The changes of field strength and section are 
important, but they have been included in the
formalism among the magnetic 
amplification de-amplification mechanisms.
We are left with changes of magnetic field direction,
in particular, with  changes of magnetic field 
inclination since the PDF is immune to the
azimuths; see equations~(\ref{rectangle2}) 
and (\ref{area}). We assume, however, that the random 
granular motions do not change inclinations in a preferred 
direction which, together with equations~(\ref{rectangle2}) 
and (\ref{area}), lead to,
\begin{equation}
{{dP_4(B)}\over{dt}}\simeq 0.
\label{buffeting}
\end{equation} 
Some limitations of this assumption are 
pointed out in \S~\ref{discussion}

Inserting equations~(\ref{mess3}), (\ref{eq2}), (\ref{eq3}),
and (\ref{buffeting}) into
equation (\ref{eq0}), one ends up with the following differential
equation for the shape of $P(B)$,
\begin{equation}
{{d\big[\kappa(B)\,B\, P(B)\big]}\over{dB}}
\simeq -\beta(B)\,B\,P(B)
+\alpha(B)\,B\, M(B).
\label{maineq0}
\end{equation} 
By definition, the three coefficients that characterize the
three physical mechanisms modifying $P(B)$ are all
positive, i.e., $\alpha(B)> 0$, $\beta(B) > 0$
and $\kappa(B)> 0$. As we point out above, $\kappa(B)$
stands for the {\em speed} of the amplification mechanism
whereas $\alpha(B)$ and $\beta(B)$ represent inverse
of time scales for the vertical transport of magnetic
structures.

From now on we drop from the 
equations the dependence of all  variables
on the field strength, which provides manageable 
expressions without compromising clarity. Then 
equation~(\ref{maineq0}) becomes,
\begin{equation}
{{d(\kappa\, B\, P)}\over{dB}}=-\beta\, B\, P+\alpha\, B\, M.
\label{maineq}
\end{equation}
It is a first order linear differential equation and,
consequently, 
its solutions are given by,
\begin{equation}
P=(B\,\kappa)^{-1}\int_0^B\alpha\, B'\,M\,
\exp[-\int_{B'}^B(\beta/\kappa)\,dB'']\,dB',
\label{fsolution}
\end{equation}
where we have considered that the
product $\kappa\,B\,P\rightarrow 0$
when $B\rightarrow 0$.
Note that the two coefficients $\alpha$ and $\beta$ cannot
be independent since $P$ is a PDF and therefore it must
be properly normalized,
\begin{equation}
\int_0^\infty P\,dB=1.
\label{normalization}
\end{equation}
According to equation~(\ref{fsolution}), $P$ scales
with $\alpha$, therefore, given $\kappa$ and $\beta$, 
a scaling factor applied to $\alpha$
guarantees the proper normalization. It follows  
from the solution of the differential equation,
\begin{equation}
{{d\big[\kappa\,B\, f\big]}\over{dB}}=
-\beta\,B\,f +\alpha_f\,B\, M,
\end{equation} 
which is identical to the original equation~(\ref{maineq})
except that $\alpha$ has been replaced with an unconstrained
$\alpha_f$. One can readily show by 
direct substitution that, 
\begin{equation}
P=f/\int_0^\infty f\, dB,
\end{equation} 
is a properly normalized
solution of equation~(\ref{maineq}) with,
\begin{equation}
\alpha=\alpha_f/\int_0^\infty f\, dB.
\end{equation}

\section{Properties of the solutions}\label{properties}

\subsection{Hump at large field strength}\label{humpit}
First and most important in the context of this paper,
equation~(\ref{maineq}) predicts PDFs with a hump
at large magnetic field strengths. The only condition
is an abrupt cease of the magnetic amplification mechanism 
when $B$ approaches a maximum value $B_{\rm max}$.
Explicitly, 
\begin{equation}
\kappa\rightarrow 0~~{\rm when}~~B\rightarrow B_{\rm max},
\label{conditiona}
\end{equation}
with 
\begin{equation}
\big|{{d\kappa}\over{dB}}\big|~~~{\rm large~enough}.
\label{conditionb}
\end{equation}
Equation~(\ref{conditiona}) puts forward a very natural  
condition to be satisfied by any amplification 
mechanism. The magnetic structures must
be in mechanical balance within the photosphere and,
therefore, the magnetic field strength cannot provide
a magnetic pressure exceeding  the (gas) pressure
of the quiet photosphere \citep[e.g., ][]{spr81b}. 
At the base of the photosphere and in the standard 1D model
atmospheres \citep[e.g.][]{mal86},
\begin{equation}
B_{\rm max}\simeq 1800~{\rm G}.
\end{equation}
In order to prove that conditions (\ref{conditiona}) 
and (\ref{conditionb}) necessarily produce a hump,
one rewrites 
equation~(\ref{maineq}) as,
\begin{equation}
{{dP}\over{dB}}\simeq 
-\big[{{\beta+{{d\kappa}/{dB}}}\over{\kappa}}+B^{-1}\big] P,
\label{maineq1}
\end{equation} 
where we have assumed that the strong fields are produced
by the amplification mechanism rather that transported
upward by the granular flows ($M\simeq 0$ when
$B\rightarrow B_{\rm max}$).
Consider the condition~(\ref{conditiona}). It forces 
\begin{equation}
P\rightarrow 0~~{\rm when}~~B\rightarrow B_{\rm max},
\label{limit}
\end{equation}
since there are no sources ($M\simeq 0$)
and the magnetic fields cannot
be intensified to $B=B_{\rm max}$. 
Because of the limit~(\ref{limit}) and the constraint $P > 0$,
\begin{equation}
{{dP}\over{dB}} < 0 ~{\rm when}~
 B\rightarrow B_{\rm max}.
\end{equation}
Consequently, the condition for a continuous
$P$ to have a maximum 
($dP/dB=0$) is equivalent to
\begin{equation}
{{dP}\over{dB}} > 0,
\label{eqabove}
\end{equation}
at a $B < B_{\rm max}$.  Using 
equations~(\ref{maineq1}) and (\ref{eqabove}),  
there is a maximum if
\begin{equation}
{{d\kappa}\over{dB}}<  
-\big[\beta+{{\kappa}\over{B}}\big],
\end{equation}
or 
\begin{equation}
\big|{{d\kappa}\over{dB}}\big|> 
\big[\beta+{{\kappa}\over{B}}\big].
\end{equation}
These two equations provide specific expressions for 
what {\em large enough} means in equation~(\ref{conditionb}).

\subsection{PDF for small magnetic field strengths}\label{rayleigh}

We expect the direction of the magnetic field vector 
to be random when $B\rightarrow 0$. In this case
the magnetic forces cannot back-react on the flows and
the magnetized plasma is freely dragged, bended and 
moved around  by the granular motions. This continuous buffeting 
impinges a random tilt to the magnetic fields,  
which should show no preferred orientation.
\citet{dom06} argue that this random orientation 
forces,
\begin{equation}
P\rightarrow 0~~{\rm when}~~B\rightarrow 0.
\label{limit0}
\end{equation}
For a point in the atmosphere to have $B=0$, the three 
Cartesian components of the magnetic field vector
must be zero simultaneously,
and this is a very improbable event when the orientation
of the field is random and therefore the three components
independent.
Here we go a step further and conjecture that $P$ should
follow a Maxwellian distribution, which is 
the PDF characteristic of the modulus of a vector field
with random orientation \citep[see, e.g.,][]{pro90,agu81}.
Explicitly, 
\begin{equation}
P\simeq {{4}\over{\sqrt{\pi}\,\sigma^3}}\,B^2\,\exp(-B^2/\sigma^2),
\end{equation} 
when $B\rightarrow 0$. 
In order to obtain this particular shape, $M$
in equation~(\ref{maineq})
must follow a Rayleigh distribution,
\begin{equation}
M\simeq {2\over{\sigma_m^2}}\,B\,\exp(-B^2/\sigma_m^2),
\end{equation}
with $\sigma=\sigma_m\sqrt{5/3}$ and
$\alpha=18\kappa/5\sigma\sqrt{\pi}$.
This can be checked by direct substitution 
of $M$ and $P$ into equation~(\ref{maineq}),
considering that $\kappa$ and $\alpha$ do not
vary with the magnetic field when $B\rightarrow 0$,
and neglecting the influence of the 
magnetic flux transported downward
($\beta P\ll \alpha M$ when $B\rightarrow 0$). 

\subsection{Analytical approximation}\label{analytical}
It was found by \citet{dom06} that the PDFs produced by 
numerical simulations of magneto-convection follow a 
lognormal distribution, and this shape was adopted to describe
$P$ for sub-kG magnetic field strengths.
In this case,
\begin{equation}
P= {{1}\over{\sqrt{\pi}\,s\,B}} \exp\big\{-\big[{{\ln (B/B_s)}\over{s}}\big]^2\big\},
\label{lognormal}
\end{equation}
with the parameters $B_s$ and $s$ related to the mean and the
variance of the distribution.
It turns out that equation~(\ref{maineq}) predicts a 
lognormal distribution when (a)
the influence of $M$ is negligible (e.g., when B is
larger than the typical field strengths supplied by the
granular upflows), and (b)
the various coefficients
characterizing the differential equation
have a weak dependence on the field strength. 
Equation~(\ref{maineq}) can  be rewritten as,
\begin{equation}
{{d\ln(B\,P)}\over{d\ln B}}=-\big[
{{\beta}\over{\kappa}}B+{{d\ln\kappa}\over{d\ln B}}\big].
\label{maineq_this}
\end{equation}
When the variation of the right-hand-side term of this
equation with magnetic field is not 
is not very large, then it can be approximated
as, 
\begin{equation}
{{\beta}\over{\kappa}}B+{{d\ln\kappa}\over{d\ln B}}
\simeq a+b\ln B,
\label{approx_logn}
\end{equation}
with $a$ and $b$ two constants\footnote{
Consider a field strength $B_1$ typical
of the range of $B$ where the approximation holds. 
Since $M$ has to be negligible, $B_1$ is
well above zero and $B$ can be expanded
as a polynomial of $\ln(B/B_1)$, namely,
$B/B_1=\exp[\ln(B/B_1)]=1+\ln(B/B_1)+[\ln(B/B_1)]^2/2+
\dots$ By construction $B/B_1\sim 1$ 
and,  therefore, the second and higher order terms of the
expansion can be neglected, which 
justifies using a logarithm to approximate $B$ in 
equation~(\ref{approx_logn}).
}. 
In this case the integration
of equation~(\ref{maineq_this}) automatically gives
a lognormal (\ref{lognormal}) with,
\begin{eqnarray}
\ln B_s=&-a/b,\cr
s=&\sqrt{2/b}.
\end{eqnarray}
Therefore, provided that the
approximation~(\ref{approx_logn}) is valid, a lognormal 
yields a good analytical representation of the
PDFs proposed in the paper. 

\subsection{Magnetic flux and magnetic energy}\label{conservation}

The two terms on the right-hand-side of equation~(\ref{maineq}) 
represent the rate of magnetic flux
emergence and submergence (\S~\ref{setup}). Since the magnetic 
amplification mechanisms
do not change the flux (\S~\ref{setup}), they have to be equal
once integrated over the full range of field strengths. This 
property is
automatically fulfilled by the solutions of equation~(\ref{maineq}).
Integrating~(\ref{maineq}) from
$B=0$ to $B=B_{\rm max}$ and considering the limit~(\ref{limit}), then,
\begin{equation}
\int_0^{B_{\rm max}}\,\beta\, B\,P\,dB=
\int_0^{B_{\rm max}}\,\alpha\, B\, M dB,
\end{equation} 
as demanded by the conservation of magnetic flux.
The amplification mechanisms transform weak fields into strong 
fields. The energy of the  distribution scales with the square 
of the magnetic field strength, therefore, the magnetic 
amplification mechanism is expected to create magnetic energy. 
One can show that any amplification mechanism increases
the energy coming from below
multiplying equation~(\ref{maineq}) by $B$ an integrating
from $B=0$ to $B=B_{\rm max}$. A trivial manipulation
leads to,
\begin{equation}
\int_0^{B_{\rm max}}\,\beta\, B^2\,P\,dB=
\int_0^{B_{\rm max}}\,\alpha\, B^2\, M dB
+\int_0^{B_{\rm max}}\,\kappa\, B\, P dB.
\label{energy}
\end{equation}
Keeping in mind that the second term of the right-hand-side
is always positive ($\kappa$, $B$ and $P$ are positive),
then
\begin{equation}
\int_0^{B_{\rm max}}\,\beta\, B^2\,P\,dB >
\int_0^{B_{\rm max}}\,\alpha\, B^2\, M dB.
\end{equation}
This inequality implies that magnetic energy transported
downward (proportional to the left-hand-side term) is always larger
than the energy coming from below (the right-hand-side term).

\subsection{Examples of PDFs}\label{examples}

\begin{figure}
\includegraphics[width=0.45\textwidth]{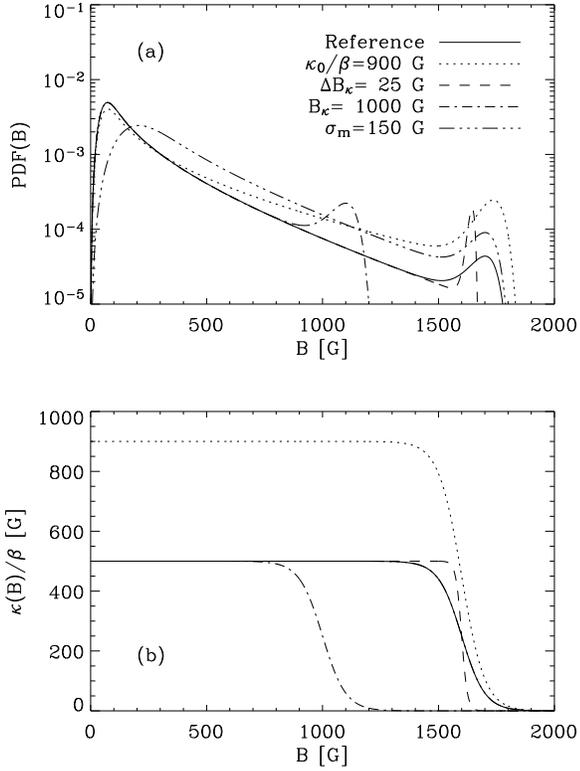}
\caption{(a) PDFs predicted by equation~(\ref{maineq})
depending on the speed of the amplification mechanism 
$(\kappa)$. 
(b) Speeds responsible for the above PDFs.
The legend in (a) indicates the only parameter
that differs between the {\em reference}
(the solid line) and the other PDFs. 
}
\label{pdf0}
\end{figure}
In order to illustrate some of the
properties described above, we integrate
equation~(\ref{maineq}) under simple assumptions
that still provide a hump at large field 
strengths. We consider  $\alpha$ and $\beta$
to be constant, whereas $\kappa$ is almost constant
up to a magnetic field $B_\kappa$ where it abruptly
drops to zero in an interval $\Delta B_\kappa$,
namely,
\begin{equation}
\kappa={{\kappa_0}\over{2}}\big[1-\tanh
	\big({{B-B_\kappa}\over{\Delta B_\kappa}}\big)\big]. 
\label{mykappa}
\end{equation}  
The symbol $\kappa_0$ stands for  $\kappa$
when $B\ll B_k-\Delta B_k$.
The differential equation is integrated using a standard 
fourth-order Runge-Kutta method, from low to strong fields, 
and using as boundary condition that given in 
equation~(\ref{limit0}). The feeding function $M$ is 
taken to be a Rayleigh distribution as discussed 
in \S~\ref{rayleigh}.
The solid line in Figure~\ref{pdf0}a shows $P$
for $B_\kappa=1600~$G, $\Delta B_\kappa=$100~G,
$\kappa_0/\beta=$500~G and $\sigma_m=50$~G,
and it is used as a reference to explore the
dependence of $P$ on $\kappa$. The curves
in Figure~\ref{pdf0}a have been computed using
the speeds shown in  Figure~\ref{pdf0}b.
First, note how all PDFs show a terminal
hump. The hump rises as the jump of $k$
sharpens (see the dashed line, with
$\Delta B_\kappa=$25~G), and its position
moves according to
$B_\kappa$ (see the dotted-dashed line, with
$B_\kappa=$1000~G). The faster the magnetic 
amplification mechanism the larger the tail of strong 
fields. The dotted line corresponds to a 
$\kappa_0$ almost twice the reference value.
Finally, the tail of strong fields also increases   
with the magnetic flux supplied by the granulation upflows.
The triple dot-dash line is fed with three times
more magnetic flux than the reference PDF. This may explain why 
plage regions have more  kG fields than the 
quiet Sun.

\section{Inversion of selected PDFs}\label{selected}

Knowing $P$ it is possible to derive by inversion 
the function $\kappa$ characterizing the concentration 
mechanism. This is particularly simple for the range
of large field strengths, where one expects $M$ to be 
negligible. Under this condition, equation~(\ref{maineq}) 
admits a  solution of the kind,
\begin{equation}
{{\kappa}\over{\beta}}
\,B\,P=\int_{B_0}^{B_{\rm max}}\,B'\,P\,dB'-
\int_{B_0}^{B}\,B'\,P\,dB'.
\label{inversion}
\end{equation}
The symbol $B_0$ stands for the smallest field strength where 
$M$ can be  neglected ($M\simeq 0$ for $B > B_0$). The coefficient 
$\beta$ has been assumed 
to be constant for convenience, but this condition
can be relaxed if required. 
Equation~(\ref{inversion}) provides  $\kappa$
given $P$.
It has been applied to a set of PDFs to illustrate 
the feasibility of the inversion. 
Theses PDFs are shown in Figure~\ref{pdf}a, whereas
the associated $\kappa$ are in Figure~\ref{pdf}b. 
The solid line corresponds to a
semi-empirical PDF that reproduces the level of
Zeeman and Hanle signals observed in the quiet Sun 
\citep[][ the {\em reference} PDF]{dom06}. Despite all the 
uncertainties, it represents the kind of the
PDF to be expected in the quiet Sun. The inversion
provides $\kappa/\beta$, i.e., the speed of the 
concentration mechanism times the time-scale for the 
granular downdrafts to remove all magnetic 
structures 
from observable layers. In order to reproduce the
semi-empirical PDF, $\kappa/\beta$ has to be of the order
of 1000~G (between 500~G and 1700~G, according to the
solid line in  Fig.~\ref{pdf}b). In other words, 
during a time scale 
$\beta^{-1}$, the concentration mechanism must 
increase the field strength from zero to 1kG. 
If we associate $\beta^{-1}$ with the granulation 
turn-over time, the 
concentration mechanism must operate with a time-scale
of, say, 15~min. 
This value is not very different from the 
time-scales characteristic of the concentration mechanisms
proposed in the literature. For example,
the convective collapse simulations by \citet{tak99}
create kG fields in minutes, whereas the thermal
relaxation proposed by \citet{san01c} easily 
concentrates low magnetic flux structures
within a granulation turn-over time. 
The other PDFs in Figure~\ref{pdf}a correspond to
numerical simulations of magneto convection by 
\citet{vog03b} having three different levels of unsigned
magnetic flux (see the inset). The values of $\kappa/\beta$
are similar to that of the semi-empirical PDF, 
however, the variation with the field strength is significantly
different. They peak  at hG field strengths 
whereas the semi-empirical $\kappa$ has its peak at 
kG field strengths. The main effect is a reduction
of the kG fields of the numerical PDFs with respect 
to the semi-empirical PDF. As a by product, the 
gradient of $\kappa$ with $B$ is smaller in the numerical
PDF which, according to the discussion in 
\S~\ref{humpit}, necessarily produces humps that are 
smaller.

\section{Discussion and conclusions}\label{discussion}

We derive a first order linear differential equation
describing the shape of the probability density
function  of  magnetic field strengths in 
the quiet Sun (PDF).
The modeling  considers  convective motions which 
continuously supply 
and remove magnetic structures from the observable
layers. In addition, a generic magnetic 
amplification mechanism tends to increase the field 
strength up to a maximum threshold.
The basic ingredients that we consider
yield PDFs in good agreement 
with the PDFs produced by
realistic numerical simulations of magneto convection,
as well as with quiet Sun PDFs inferred from observations.
In particular, they give a natural explanation for 
the hump right before the cutoff at 1700~G found by 
\citet{dom06b,dom06}. It is produced by the abrupt end
of the amplification mechanism at the cutoff. 
The solutions
of our equation provide a  
parametric family of PDFs  as required
for diagnostics in automatic inversion 
procedures \citep[see][]{soc01}. 
For example, the PDFs analyzed in \S~\ref{selected} depend
on four free parameters which, in principle, can be
fitted using four independent observables.
We also derive an approximate expression to infer the speed of 
the amplification mechanism once the PDFs are known.
In order to reproduce the semi-empirical
PDFs, the amplification
mechanism must concentrate field strengths
from zero to 1~kG in a time-scale similar
to the granulation turn-over time. 

The magnetic structure of the quiet
photosphere is certainly much more complex
than the modeling carried out
in the paper, as it is evidenced by the existing 
numerical simulations of magneto convection 
(see \S~\ref{introduction}).
To mention just a few potentially important
ingredients that we have ignored,
the speed of the amplification 
may depend on the section of the magnetic 
structure \citep{spr79}, we expect a tendency
for the plasma with large field strength to be buoyant 
and so vertical \citep[e.g.][]{sch86}, and
the fluxtubes forming loops fully embedded in
the photosphere do not change their volume 
under magnetic amplification. 
This caveat notwithstanding, one should not 
underestimate the ability of the model to 
reproduce realistic PDFs. 
It is a non-trivial property which may reflect that,
even if sketchy and primitive, the model contains
the basic physical ingredients shaping the 
PDF of the quiet Sun magnetic fields.
Obviously, we cannot discard that the agreement is
due to a fortunate coincidence, and so
the model is offered here only as a mere possibility to 
interpret the quiet Sun PDF.

%

%
%
%

\acknowledgements

Thanks are due to F.~Kneer and I. Dominguez Cerde\~na for 
stimulating discussions on the quiet Sun magnetism, and
to A.~V\"ogler for providing the PDFs shown in Figure~\ref{pdf}.
The work has partly been funded by the Spanish Ministry 
of Education and Science, project AYA2004-05792.
%
%
%

%

\end{document}